\documentclass[twocolumn, twocolappendix]{aastex631}

\usepackage[mathscr]{euscript}
\usepackage{comment}
\usepackage{hyperref}
\usepackage{amsmath}
\usepackage[T1]{fontenc}

\definecolor{indigo}{rgb}{0.0, 0.25, 0.42}
\definecolor{forestgreen}{rgb}{0.13, 0.55, 0.13}

\begin{document}

\title{An Introduction to Dust Evolution and Vertical Transport in Protoplanetary Disks}

\email{marion.villenave@univ-grenoble-alpes.fr}
\author[0000-0002-8962-448X]{Marion Villenave}
\affiliation{Univ. Grenoble Alpes, CNRS, IPAG, F-38000 Grenoble, France}
 
\begin{abstract}
This tutorial is an introduction to observational studies of dust transport and evolution in protoplanetary disks. Spatially resolved observations of disks at multiple wavelengths can allow to infer the distribution of various dust grains and gas species. Combining these observations offers a more complete understanding of dust structure and properties across different disk locations. 
For example, by better characterizing the disk vertical structure, observations help to constrain the level of vertical settling and identify regions of high dust density, which are favorable for grain growth and planet formation. This tutorial describes various methodologies for inferring dust properties and vertical height of different tracers, as an introduction for beginners.

\end{abstract}

\section{Introduction} \label{sec:intro}

Planets form around young stars from small dust particles present in the interstellar medium. As the star gets built, the initial surrounding envelope of gas and dust evolves into a disk. In the protoplanetary disk, also called the Class~II phase of star formation~\citep[e.g.,][]{Lada_1987}, the material is denser than  in the previous envelope phases, refereed to as Class~0 and Class~I. This allows the initially submicron sized particles to collide more frequently and aggregate to larger bodies at a faster rate than in the previous stages. In the core accretion scenario, the progressive aggregation of dust to bigger sizes leads to the formation of kilometer sized bodies, also called planetesimals, and rocky planetary cores. Eventually, those cores can become sufficiently massive to accrete the surrounding gas and become giant planets. The formation and growth of planetesimals are accelerated by mechanisms such as the streaming instability~\citep{Youdin_2005, Johansen_Youdin_2007, Johansen_2014} and pebble accretion~\citep{Ormel_2010, Lambrechts_2012}, which necessitate a significant accumulation of dust particles to become efficient. 

In a protoplanetary disk, however, dust is  not only susceptible to aggregate and grow. Instead, if the collision velocities are too large, dust grains can also fragment to smaller sizes, or bounce without growing or fragmenting~{\citep[][]{blum_2008, Wada_2011, Wurm_2021, Hasegawa_2021}}. These phenomenons are barriers to efficient grain growth.  

Besides, the interaction of dust particles with the relatively slower gas leads dust to lose angular momentum. This implies that dust particles fall vertically into the midplane, via a mechanism called vertical settling, and also drift radially toward the star or toward a local radial pressure maximum (see Sect.~\ref{sec:theorydrift}, \autoref{eq:drift}).  Vertical settling favors the accumulation of dust in the midplane, while radial drift depletes it from some regions. 
To understand how planets form, it is thus important to characterize the spatial distribution of dust, in order to identify regions of increased density and potentially more efficient growth. This is possible using spatially resolved observations of disks at different wavelengths.

In this tutorial, we first review theoretical concepts on dust transport, in particular vertical settling and radial drift~(Sect.~\ref{sec:theorydrift}), and on grain growth or fragmentation~(Sect.~\ref{sec:graingrowth}). 
In Sect.~\ref{sec:multiwavelength}, we highlight how observations at multiple wavelengths trace grains of different sizes and at different temperatures.  Sect.~\ref{sec:SEDanalysis} focuses on the description of a method which uses radially resolved observations at multiple wavelengths in the millimeter range to infer {local} disk and grain properties. Then, Sect.~\ref{sec:settling} details different methods used to infer constraints on dust transport in disks. We specifically focus on constraints on the vertical distribution of dust and gas in protoplanetary disks. Because inclination is a critical parameter for the applicability of the various methods, we differentiate edge-on disks from moderate inclination systems. Finally, Sect.~\ref{sec:conclusion} presents a summary of the different sections. 

\section{Basics of dust transport and evolution}

In this section, we summarize the key theoretical concepts related to dust evolution mechanisms. Sect.~\ref{sec:theorydrift} discusses dust transport, specifically vertical settling and radial drift, while Sect.~\ref{sec:graingrowth} presents key concepts related to grain and planetesimal growth. For a more detailed theoretical description, we refer the interested reader to the recent review of \cite{Birnstiel_2024}.

\subsection{Vertical settling and radial drift}
\label{sec:theorydrift}

Gas in a protoplanetary disk experiences both the stellar gravity and its own pressure. This implies that gas orbits the star at sub-Keplerian velocities. On the other hand, without the presence of the gas, dust particles would only experience the stellar gravity and orbit the star  at exactly Keplerian speed. In the presence of gas, dust particles feel a head-wind, or a drag, coming from the slower gas which affects their orbits~\citep[e.g.,][]{Weidenschilling_1977, Nakagawa_1986}. A dust particle, with an initial inclination and eccentricity with respect to the gas orbit, will see its vertical and radial oscillations with respect to the gas rotation damped by the drag forces until its orbit corresponds to that of the gas. Because dust typically accounts for only 1\% of the gas mass, dust particles usually experience a much larger change in velocity than the gas, even if the energy lost by the dust is transferred to the gas.

The impact of gas on dust particles depends on how well they are coupled. The dust particles which can be directly observable in protoplanetary disks (see Sect.~\ref{sec:multiwavelength}) are typically small compared to the mean free path of a gas molecule. This is the co-called Epstein regime. In that case, the energy transfer between gas and dust is caused by molecular collisions, with a typical timescale, also called stopping time, of:
\begin{equation}
    \tau_\text{s} \propto \frac{a}{c_\text{s}}\frac{\rho_\text{s}}{\rho_\text{g}}
    \label{eq:tauS}
\end{equation}
where $a$ is the particle size,  $\rho_\text{s}$ the density of a dust particle, and $\rho_g$ the gas density in the disk. $c_\text{s} = \sqrt{\frac{k_\text{B}T}{\mu m_\text{p}}} $ is the sound speed, which depends on the Boltzmann constant $k_\text{B}$, the local disk temperature $T$, the mean molecular weight $\mu\sim2.3$, and the mass of a proton $m_\text{p}$. 

For completeness, we also note that the gas density can be re-written as $\rho_\text{g} = \Sigma_\text{g} / H_\text{g}$, where $\Sigma_\text{g}$ is the gas surface density and $H_\text{g}$ the gas scale height. At the hydrostatic equilibrium, the gas scale height characterizes the vertical extension of the gas, which vertical distribution follows $\rho(z)\propto e^{-z^2/2H_\text{g}^2}$. For a vertically isothermal and non self-gravitating disk, the gas scale height can be estimated by:
\begin{equation}
        H_\text{g}(r) = \frac{c_\text{s}}{ \Omega_\text{K} }= \sqrt{\frac{k_\text{B}Tr^3}{GM_\star\mu m_\text{p}}}
    \label{eq:gasScaleHeight}
\end{equation}
where  $ \Omega_\text{K} = \sqrt{GM_\star / r^3}$ is the angular Keplerian velocity, which depends on the radius $r$, the gravitational constant $G$, and the stellar mass $M_\star$.

Because the stopping timescale  ($\tau_\text{s}$, \autoref{eq:tauS}) can vary a lot within one disk, it is usually compared with the typical orbital timescale of the disk at a given radius, given by the angular Keplerian velocity $\Omega_K $. This dimensionless ratio is called the Stokes number. In the Epstein regime, it can be written as:
\begin{equation}
    \text{St} =  \Omega_\text{K} \tau_\text{s} = \frac{a \rho_\text{s}}{\Sigma_\text{g}}\frac{\pi}{2}
    \label{eq:St}
\end{equation}

Small Stokes number ($\text{St}\ll1$) are well coupled to the gas, while particles with a larger Stokes number will be increasingly decoupled.
In~\autoref{eq:St}, we can see that, in the common Epstein regime, the coupling of dust and gas  linearly depends on the size of the particle $a$ and on the gas surface density $\Sigma_\text{g}$. For the same gas density, smaller particles are better mixed with the gas than larger grains. On the other hand, grains of similar sizes will be more easily mixed with the gas if the gas density is higher. \\

Within this framework, dust is expected to migrate radially and vertically with the following velocities~\citep[see e.g.,][for more details on the derivations]{Birnstiel_2024}:
\begin{subequations}
    \begin{align}
     &v_{z\text{, settling}}= -z \Omega_\text{K} \text{St} \label{eq:settl}\\
       &v_{r\text{, drift}} = \frac{v_\text{K}}{\text{St} + \text{St}^{-1}} \left(\frac{H_\text{g}}{r}\right)^2 \frac{\partial ln P}{\partial ln r} \label{eq:drift}
    \end{align}
\end{subequations}

where $v_\text{K} =\sqrt{GM_\star / r}$ is the Keplerian velocity, $P$ the gas pressure, and $z$ the height above the midplane. {\autoref{eq:settl} and \ref{eq:drift} are} valid if the dust-to-gas ratio is small ($\ll1$).

Using these equations, we can see that vertical settling speed is faster for large Stokes number (i.e., for large dust particles or small gas density). Moreover, because the settling velocity becomes null at $z=0$, {\autoref{eq:settl}} shows that grains are settling down to the midplane. 

On the other hand, in the radial direction, one can easily see that grains of Stokes number close to 1 are drifting at the highest velocity. Smaller grains remain well coupled with the gas, while larger bodies are not affected by the gas. {\autoref{eq:drift}} shows that grains drift following the pressure gradient, towards a maximum of pressure. The sign of the radial drift speed changes with the sign of the pressure gradient. A local inversion of the pressure gradient can thus lead to a change in the drag speed, pushing grains from both sides towards the pressure maximum. If such gradient is stable in time, it can efficiently trap dust particles.\\

These considerations are valid for gas which is only subject to the stellar gravity and its own pressure. However, the gas might also be turbulent~\citep[see][for a recent review]{Rosotti_2020}, which leads to more complex gas velocities.  {Various mechanisms are thought to be able to generate turbulence, depending on the disk location and local conditions~\citep[see e.g.,][for a review]{Lesur_2023}, leading to different predictions for the strength and morphology of turbulence. For example,   the isotropic magneto-rotational instability~\citep[MRI,][]{Balbus_1991} is primarily active in highly ionized environments, such as the very inner disk or the outer disk ($R\lesssim1$\,au and $R\gtrsim100$\,au). Conversely, hydrodynamic instabilities, such as the vertical shear instability~\citep[VSI,][]{Nelson_2013} which leads to strong vertical but weak radial turbulence, likely occurs in less ionized areas. Gravitational instability~\citep[GI,][]{Goldreich_1973} can manifest itself on large scales within cold and massive disk regions. This mechanism leads to stronger radial than vertical turbulence. Beyond these well studied mechanisms, other potential drivers of turbulence include, among others, the streaming instability and embedded planets.} 

The presence of turbulence leads to additional mixing on the particles and {typically} acts as a barrier for efficient settling or radial drift {(although see \citealt{Gerosa_2023})}. Specifically, assuming a balance between turbulent spreading and the expected settling,  and for low Stokes number ($\text{St}\ll1$), the vertical dust-to-gas concentration follows~\citep[e.g.,][]{Youdin_2007, Dubrulle_1995}:
\begin{equation}
       \frac{H_\text{d}}{H_\text{g}} = \left( 1 + \frac{\text{St}\text{Sc}}{\alpha_z}\right)^{-1}
    \label{eq:alphaZ_St}
\end{equation}
where $H_\text{d}$ is the dust scale height and $\alpha_z$ the level of vertical turbulence. Sc is the Schmidt number, which quantifies the ratio between gas and particles diffusivity. It is often approximated to unity~\citep[e.g.,][]{Fromang_Nelson_2009, Pinte_2016, Dullemond_2018}.

In addition, if there is a radial pressure bump in the disk, dust will be trapped at the pressure maximum, creating rings. In that situation, the dust width $w_d$ can be related to the gas width $w_g$ following a very similar equation~\citep[e.g.,][]{Dullemond_2018}: 
\begin{equation}
    \frac{w_\text{d}}{w_\text{g}} = \left( 1 + \frac{\text{StSc}}{\alpha_r}\right)^{-1}
\end{equation}
where $\alpha_r$ is now the turbulence along the radial direction.

Characterizing how settled dust is compared to the gas provides constraints to the vertical level of turbulence. On the other hand, measuring the relative radial width of dust and gas rings can provide insights towards the radial turbulence, and whether dust is trapped in a pressure maximum. Better characterizing the turbulence structure and level in disks can allow to access further insights towards its physical origin because different mechanisms predict different turbulence strength and morphology~\citep{Lesur_2023}. 
Besides, characterizing the degree of dust concentration will allow to reveal its potential for efficient growth, as we describe in the next section.

\subsection{Grain size evolution}
\label{sec:graingrowth}

Dust properties in the interstellar medium are well characterized. Dust grains are likely a mixture of silicates and carbons, with a size distribution from $\sim\,0.01\,\mu$m to a maximum of a few microns~\citep[e.g., review by][]{Natta_2007}. Dust grains of the interstellar medium follow a size distribution such that $n(a)da \propto a^{-3.5}da$~\citep{Mathis_1977}, where $a$ is the particle size and $n(a)$ the number of particles of this size. 

During the protoplanetary disk phase, the grain sizes evolve, eventually leading to the formation of planets. These processes have been the focus of various recent reviews, which discuss many stages of the planet formation processes, from the growth of micron sized dust, to the size increase of planetary bodies~\citep[e.g.,][]{ Wurm_2021, Drazkowska_2023, Birnstiel_2024}. 
Because of the extensive pre-existing review work available, we only include the necessary arguments needed to appreciate observational constraints on the dust sizes~(Sect.~\ref{sec:SEDanalysis}). \\

The combination of vertical settling, radial drift (see Sect.~\ref{sec:theorydrift}), brownian, and turbulent motion induces large differential vertical and radial motion of dust particles. Those lead to frequent dust-dust collisions, which drive grain growth or dust size decrease. Apart from the dust composition {and shape}, the outcome of a collision depends on the dust sizes and the relative velocity~{\citep[e.g.,][]{ Wada_2009, Testi_2014, Birnstiel_2024}}. In short, collisions can lead to sticking, growth by mass transfer, bouncing, fragmentation, and erosion. Low speed collisions of particles of different (small) sizes preferentially lead to grain growth outcomes. 
In the coming paragraphs, we highlight instead some growth barriers, which are fragmentation, bouncing, and radial drift.

When the relative velocities are sufficiently large, colliding dust grains will fragment to smaller sizes rather than growing bigger. Because, the relative velocities tend to increase for larger grain sizes, we expect the grain sizes to be limited by fragmentation, bouncing,  or erosion~\citep[see e.g., figure 3 of][]{Zsom_2010}. 
Fragmentation limited dust distributions are predicted to maintain a size distribution following a power-law, with an exponent which does not necessarily match that of the interstellar medium~{\citep{Birnstiel_2011, Hasegawa_2021}}.

In some cases, bouncing sets the size limit~\citep[e.g.,][]{Zsom_2010}. In this regime, grains do not fragment but they also do not grow further. 
Contrary to the fragmentation limited case, if the dust size distribution is limited by bouncing, the dust size distribution is significantly modified and can become nearly mono-disperse~{\citep{Stammler_2023, Dominik_2024, Oshiro_2025}}.

Finally, radial drift of dust can also act as a local growth barrier. Indeed, large grains, with a Stokes number closest to 1, tend to efficiently migrate toward a pressure maximum~(\autoref{eq:drift}). This would lead to the depletion of nearby regions from large grains and can set the maximum size in these regions. In the case where drift limits the dust size, the dust distributions are predicted to follow a slightly different size distribution from the interstellar medium, with a slope of 2.5, corresponding to a relatively larger quantity of bigger particles~\citep{Birnstiel_2015, Birnstiel_2024}. \\

It is to be noted that most works focus on the evolution of compact spheres, meaning that when two particles aggregate they are then treated as a compact sphere of the corresponding mass. However, aggregates are not necessarily compact. They can be porous and have lower fractal dimension, especially at the earliest stages of grain growth{~\citep[e.g.,][]{Kataoka_2013, Garcia_2020, Wurm_2021}}. Further evolutionary stages can lead to later compaction.

Porosity is important for grain evolution and dust transport because it changes their Stokes number. A more porous particle of mass $m$ will be larger than a compact grain, and thus have a larger Stokes number. For particles larger than the wavelength, the stopping time is described by the so-called Stokes regime, and depends on the grain size squared ($\tau_\text{s, Stokes}\propto a^2$). 

Studies suggest that grain grow to larger sizes when porosity is considered~\citep[e.g.,][]{Okuzumi_2012, Garcia_2020, Michoulier_2024}. {In addition, numerical simulations of collisions between particules-aggregates (ballistic particle-cluster aggregation, BPCA) or aggregate-aggregate (ballistic cluster-cluster aggregation, BCCA) show that the number of monomers included in the aggregates, the impact parameter of the collisions, and the number of monomers in contact within a particle (average coordination number) are particularly important to determine the outcome of this type of collisions~\citep[e.g.,][]{Wada_2009, Wada_2011}. Further simulations and experiments are necessary to better understand how the internal structure of the results of collisions also affect the subsequent collision outcomes.}\\

All the barriers highlighted earlier suggest that {dust particles} can hardly grow larger than pebble sizes without being fragmented or drifting radially. However, once pebbles are present in the disk, other mechanisms can occur, allowing to bypass intermediate grain sizes and directly form planetesimals from pebbles. In particular, the so-called streaming instability can allow to form 100\,km sized bodies very quickly~\citep{Youdin_2005, Johansen_2014}. This instability requires pebbles to be formed~\citep[$\text{St}\geq10^{-3}$,][]{Drazkowska_2023} and an increased dust density in the midplane. In particular, a gas-to-dust ratio close to unity is typically required to trigger this instability~\citep{Carrera_2015, Carrera_2025, Lim_2025}. Further, planetesimals can grow via planetesimal collisions or via the accretion of pebbles, which Stokes number can be between $10^{-3}$ and 1. Details on these mechanisms have  recently been reviewed by \cite{Drazkowska_2023}. Of interest here, we emphasize that pebble accretion can be much faster than planetesimal accretion for growing a planetesimal to a planetary core. Pebble accretion is especially fast when the planetary embryo can accrete the complete layer of pebbles, which can happen if dust grains are settled into the disk midplane. 

\section{Key concepts on multi-wavelength observations}
\label{sec:multiwavelength}

\begin{figure*}
    \centering
    \includegraphics[width=\linewidth]{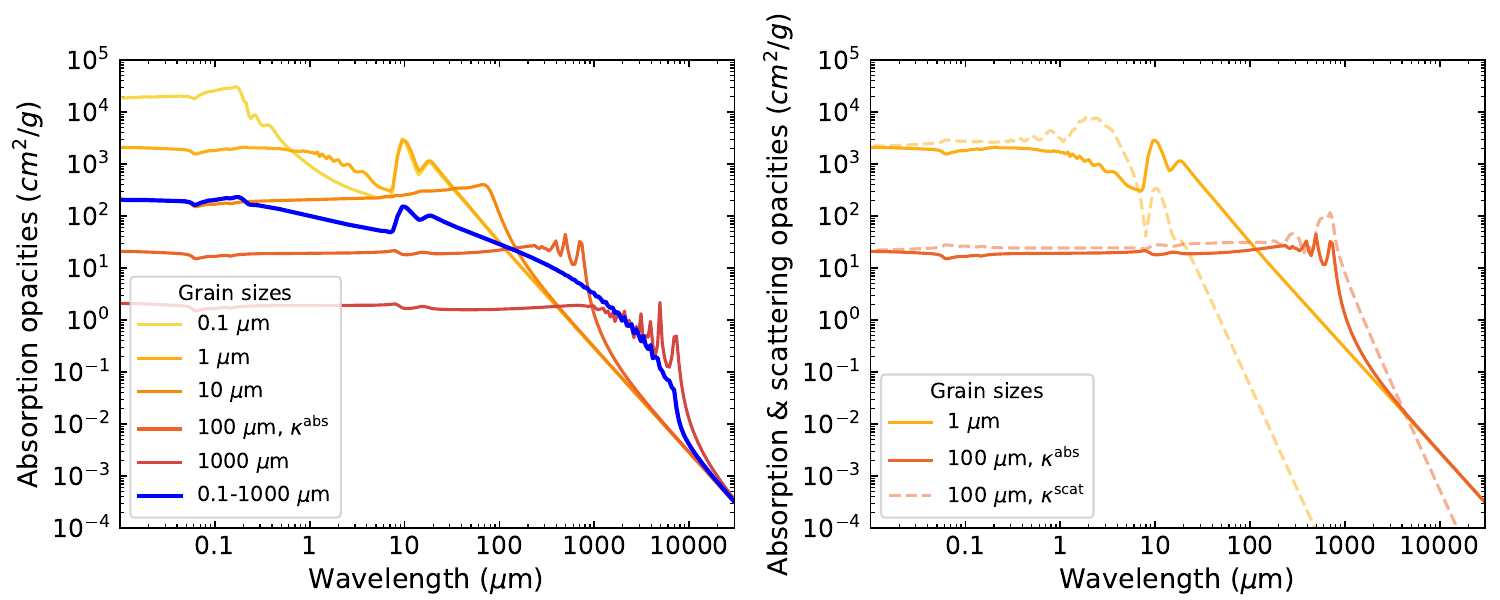}
    \caption{Opacities of astronomical silicates as a function of wavelength. {The grains are assumed to be compact spheres.} \emph{Left}: Absorption opacities of mono-disperse dust distributions (yellow to red lines)  and of a dust size distribution with grains between 0.1\,$\mu$m and 1\,mm with a number exponent of $p=3.5$ (thick blue line). \emph{Right}: Absorption (solid lines) and scattering (dashed lines) opacities for grains of 1\,$\mu$m and 100\,$\mu$m. }
    \label{fig:opacities}
\end{figure*}

Even though gas contains most of the mass of a protoplanetary disk, dust opacities dominates the continuum emission. This implies that dust dominates the outcome of the reprocessing of an incident beam light by either absorbing or scattering it. In the case of scattering, an incident beam keeps its original wavelength but changes orientation.  These effects reduce the quantity of photons traveling towards one direction and can be called dust extinction. This section discusses how multi-wavelength observations can trace various grain sizes and disk regions, thanks to different grain opacities and local disk temperatures.

\subsection{Dust opacities}
\label{sec:dustopacities}
How much dust absorbs or scatters the incident light depends on the grain properties, such as their composition, size, shape, etc. This is quantified by the dust absorption ($\kappa_\nu^\text{abs}$) or scattering ($\kappa_\nu^\text{scat}$) opacity coefficients, which are typically expressed in units of cm$^{2}$.g$^{-1}$. In Figure~\ref{fig:opacities}, we illustrate how the absorption and scattering opacities of grains of different sizes vary with wavelength. We used the dust opacities for astronomical silicates from the tables published by \cite{Laor_Draine_1993}\footnote{{The dust opacity tables can be downloaded following \href{ftp://ftp.astro.princeton.edu/draine/dust/diel/eps_Sil}{this link}}.} to illustrate the typical behavior of the opacities {of compact spherical dust particles}.\\

On the left panel of Figure~\ref{fig:opacities}, we can see that{, for compact spheres,} the absorption opacity ($\kappa^\text{abs}$) typically reaches a maximum at wavelengths around $\lambda\sim2\pi a$. For a given grain size, shorter wavelengths show relatively constant, high opacity values, while the opacity decreases significantly at {longer} wavelengths ($\lambda>2\pi a$). At the longer wavelengths, the dust opacity can be described as a power law with $\kappa^\text{abs}\propto\nu^\beta$. The power law exponent, $\beta$, is sensitive to the maximum grain size, with $\beta$ typically larger for smaller grains ($\beta\sim1.7$; and $\beta\sim0$ for mm grains). 

Moreover, at short wavelengths, one can see that the opacities of the smallest grains are significantly larger than those of larger dust particles.  When considering a full grain size population following the typical interstellar medium size variation ($n(a)da \propto a^{-3.5}da$, blue line in left panel of Figure~\ref{fig:opacities}), this implies that small grains (e.g., micron sizes) dominate the opacities at short wavelengths (e.g., up to the infrared), while large grains (e.g., sub-millimeter sizes) have an larger impact at larger wavelengths, such as millimeter wavelengths. In addition, because of the shape of the opacity curves, observations at optical to infrared wavelengths are typically more optically thick than those at longer wavelengths (e.g., millimeter wavelengths).\\

The right panel of Figure~\ref{fig:opacities} highlights how the scattering opacity compares to the absorption opacity for {compact} grains of 1\,$\mu$m and 100\,$\mu$m. We can see that the scattering curves have similar shapes to the absorption curves. For both grain sizes, the scattering and absorption opacities have similar values for wavelengths below their respective resonance ($\lambda<2\pi a$). In both cases however, scattering dominates over absorption {near} the resonance wavelength ($\lambda\sim2\pi a$). At longer wavelengths, scattering can be neglected for small grains, while it stays at comparable levels to absorption for large grains. This has implications for the interpretation of multi-wavelength millimeter observations, as we will see in more details in Sect.~\ref{sec:SEDanalysis}.

\begin{figure}
    \centering
    \includegraphics[width=\linewidth]{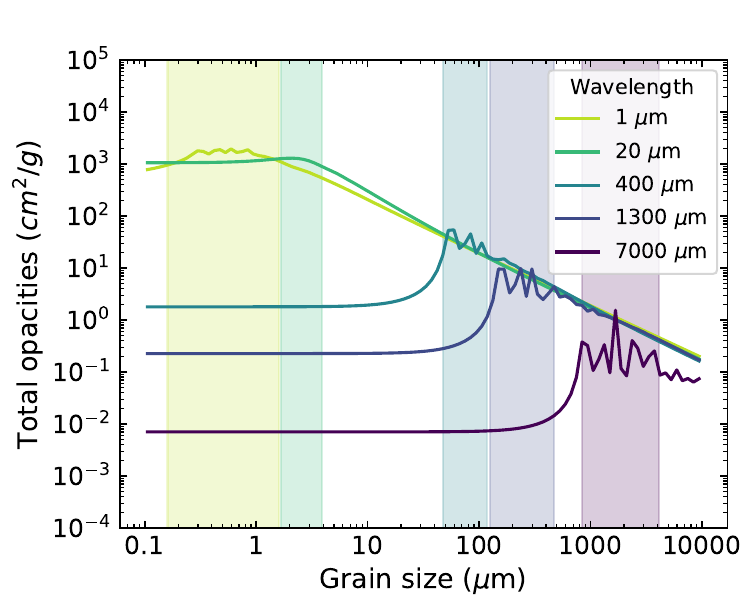}
    \caption{Total opacities as a function of grain size for different wavelengths.  {Grains are assumed to be compact spheres}. Grain sizes contributing the most at the different wavelengths are highlighted with a vertically filled region of the same color as the opacity curves. 
    }
    \label{fig:opacitieswv}
\end{figure}

To better visualize which grain sizes dominate at each wavelength {in the case of compact spheres}, we plot in Figure~\ref{fig:opacitieswv} how the total opacities (scattering + absorption) depend on grain sizes for different wavelengths. With this figure, and for the dust composition assumed here, we can see directly that emission at 1\,$\mu$m is dominated by grains between 0.1 and 1\,$\mu$m, while emission at 1.3\,mm would instead be dominated by grains between 100 and 500\,$\mu$m. Observations at widely different wavelengths are thus sensitive to grains of different sizes. \\

\begin{figure}
    \centering
    \includegraphics[width=\linewidth]{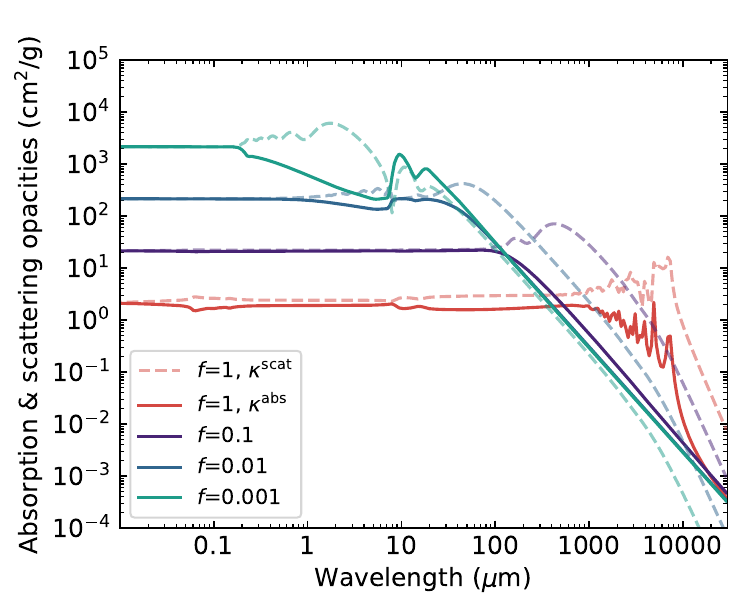}
    \caption{{Effect of porosity on the absorption (solid lines) and scattering (dashed lines) opacities of 1\,mm grains as a function of wavelength. The filling factor $f=1$ represents compact grains, while more porous particles have a lower $f$ value. } }
    \label{fig:opacities_porosity}
\end{figure}

\begin{figure*}
    \centering
    \includegraphics[width=\linewidth, trim={1cm 3cm 0cm 5cm},clip]{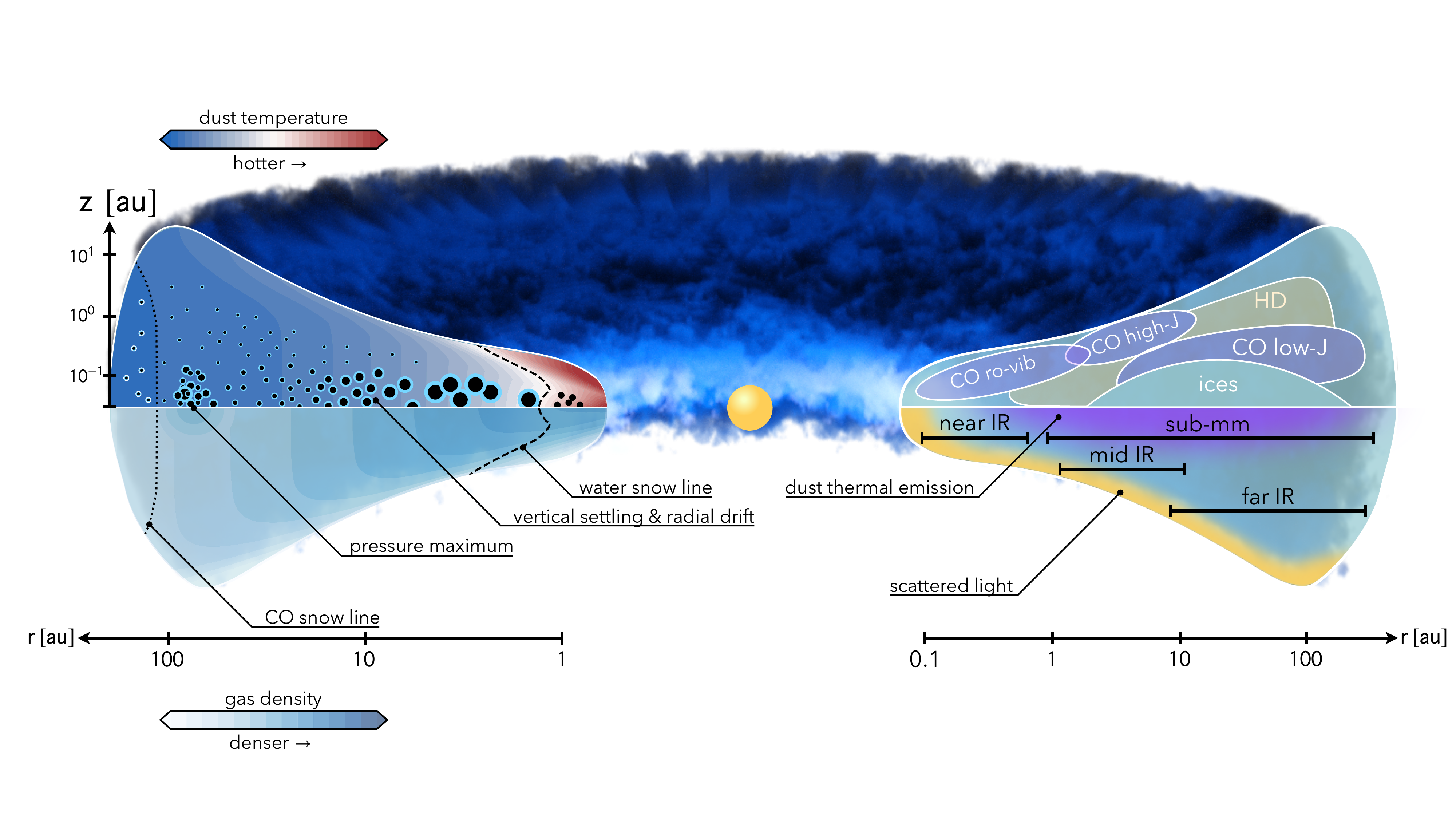}
    \caption{Illustration of the temperature structure, dust distribution, and observational constraints in protoplanetary disks from \cite{Miotello_2023}; Credit: T. Birnstiel. \emph{Left}: main dust transport mechanisms and temperature structure. \emph{Right}: typical dust and gas  emission regions.
    The axes show the logarithmic distance to the central star, in the radial and vertical direction. }
    \label{fig:diskSchematik}
\end{figure*}

{Compact spheres are simple representations of dust grains for which their optical properties can be easily calculated using the Mie theory. However, in reality, dust growth naturally forms porous aggregates (see Sect.~\ref{sec:graingrowth}), which can have very different optical properties than compact grains. Porosity is characterized by the filling factor, which is defined as the fraction of the volume of a dust particle that is occupied by matter. It quantifies the quantity of void within a dust grain. A compact grain has a filling factor equal to 1. On the other hand, more porous particles have smaller filling factors, which can be as low as $10^{-4}$ in the case of protoplanetary disks~\citep[e.g.,][]{Kataoka_2013, Garcia_2020}. }

{ In Figure~\ref{fig:opacities_porosity}, we illustrate how porosity modifies the absorption and scattering opacities of a dust particule of 1\,mm in size. We see that more porous grains of similar sizes have absorption opacities comparable to those of smaller compact grains. Specifically,  \citet{Kataoka_2014} identified that the absorption opacity is determined by the product of the grain size with the filling factor $af$ (see their Figure 2 and 3). The only difference between compact and porous grains with the same $af$ product is that the opacity increase at the resonance wavelength is removed for porous particles.  On the other hand,  \citet{Kataoka_2014} also showed that the product $af$ does not fully determine the scattering opacity. Specifically, at wavelengths longer than the resonance wavelength, the scattering opacity is typically more important for fluffier aggregates than for compact particles. This shows that porosity is an important factor to consider when interpreting dust opacities to infer dust properties. }

\subsection{Disk physical temperature}
\label{sec:temperature}
Besides absorbing and scattering incident light, dust particles can re-emit thermally the absorbed energy, or that energy can allow chemical reactions on the grain surface. We focus here on dust thermal emission and refer the interested reader to the recent review by \citet{van_Hoff_2024} for related chemical effects. 

Neglecting scattering, thermal emission of dust at the frequency $\nu$ can be written as follows: 
\begin{equation}
    I_\nu^\text{abs} = B_\nu(T_\text{dust})(1-e^{-\tau_\nu/\mu})
    \label{eq:iblackbody}
\end{equation}
with  $B_\nu(T_\text{dust})$ is the blackbody emission at the dust temperature $T_\text{dust}$, $\tau_\nu=\Sigma_\text{dust}\kappa_\nu$ is the optical depth of the dust, and  $\mu = \cos{i}$ where $i$ is the disk inclination. 
\autoref{eq:iblackbody} shows that dust emission is directly proportional to  the blackbody emission at the dust temperature. In other words, the absorbed emission re-emitted by the dust grains is distributed over the electromagnetic spectrum according to the dust temperature. 

Using this and neglecting the optical depth term for now, we can estimate the wavelength corresponding to the maximum of the emission $\lambda_\text{peak}$ in regions of different temperature $T$ using the Wien’s displacement law: {$\lambda_\text{peak} T = 2.898\times10^{-3}$ m.K}. Following this law, particles in the cold disk midplane, where $T\sim20$\,K, will have a peak of emission at long wavelengths, around 140\,$\mu$m. On the other hand, to emit thermally  $\lambda\sim1\mu$m, particles must have a temperature of about 3000\,K, which is only present in the innermost regions and upper layers of the disk. 

This highlights that, in addition to the grain sizes (see Sect.~\ref{sec:dustopacities}),  observations at different wavelengths are sensitive to emission from disk regions that have different physical temperatures. These temperature variations occur both across the disk's radius and along its vertical structure. Fig.~\ref{fig:diskSchematik} illustrates the temperature distribution in a protoplanetary disk and the different layers traced by various observing wavelengths. 

\subsection{Observing considerations}

At optical and infrared wavelengths, we expect to be sensitive to both thermal emission from the very inner disk and scattering of stellar photons by the cold outer regions. Scattering is expected to be important because at these wavelengths, micron and submicron sized particles dominate the total opacity (Figure~\ref{fig:opacitieswv}), and scattering dominates over absorption (right panel of Figure~\ref{fig:opacities}). Because of the high opacities of these particles, optical depths are high in the disk, implying that scattered photons trace the upper disk layers.  Such observations in the optical to the infrared are often referred to as scattered light observations. 

Importantly, scattering of grains which are small compared to the observing wavelength is typically not isotropic. The phase function is strongly peaked toward forward scattering{~\citep[e.g.,][]{Murakawa_2010, Min_2012, Mulders_2013, Tazaki_2019}}, which can be analyzed observationally. 

In addition, scattering induces linear polarization which is mostly perpendicular to the scattering plane defined by the central star and the observer. In other words, polarization is usually azimuthal in the disk plane, although multiple scattering and specific dust properties may also induce some radial scattering~\citep{Canovas_2015}. To isolate the disk signal from the globally unpolarized stellar emission, polarization observation are commonly used to image disks at optical and near infrared wavelengths~\citep[see][for a recent review]{Benisty_2023}.\\

On the other hand, at millimeter wavelengths, dust is optically thinner and thermal dust emission can be traced all the way to the midplane. At millimeter wavelengths, scattering can however not always be neglected{, especially if grains are porous (see Sect.~\ref{sec:dustopacities})}. Because the stellar emission is very faint, dust mostly scatters its own thermal emission at these wavelengths. This is sometimes called self-scattering, and can also lead to linear polarization at millimeter wavelengths~\citep{Kataoka_2015}. 

{Interestingly, the amount of millimeter polarized intensity can provide direct constraints towards the maximum size and porosity level of the dust particules. In simple words, compact grains scatter most efficiencly within a short range of wavelength and thus cannot generate polarization for a wide range of wavelengths~\citep[e.g.,][]{Kataoka_2016, Hull_2018}. Very fluffy aggregates on the other hand scatter efficiently over a large range of wavelengths but have a low albedo which reduces the outcoming intensity and scattering polarization. These effects can be used to obtain constraints towards the dust sizes and properties~\citep[e.g.,][]{Tazaki_2019_mm, Zhang_2023, Ueda_2024}.}

\section{Observational constraints on dust sizes}
\label{sec:SEDanalysis}
In this section we discuss observational constraints on dust grain sizes based on multi-wavelength observations. We focus specifically on the analysis of  millimeter continuum wavelengths. We refer the interested reader to the recent PPVII review of \citet{Benisty_2023} for details on constraints which can be obtained from scattered light observations, and to  \citet{Miotello_2023} for a summary of recent results based on millimeter polarization observations. Here, we aim to provide further insights on a technique recently used on high angular resolution, millimeter range, observations of disks to provide radially resolved constraints on disk and dust properties.

\subsection{Classical analysis}
\label{sec:classical_analysis}

\autoref{eq:iblackbody} has been widely used in the literature to estimate the dust properties, specifically by estimating the millimeter spectral index, $\alpha$, defined by  $I_\nu\propto\nu^\alpha$. {To interpret the millimeter spectral index, it is often assumed that these wavelengths are sensitive to dust thermal emission, and that the Rayleigh-Jeans approximation is valid. In this context, the blackbody emission follows a powerlaw, with a spectral index of 2. }

{Assuming dust thermal emission in the Rayleigh-Jeans approximation, when}
dust is optically thin ($\tau_\nu\ll1$), \autoref{eq:iblackbody} resumes to:
\begin{equation}
I_\nu\sim B_\nu(T_\text{dust})\tau_\nu\sim B_\nu(T_\text{dust})\Sigma_\text{dust}\kappa_\nu\propto \nu^{2 + \beta}
\end{equation}
where $\beta$ is the dust opacity spectral index introduced in Sect.~\ref{sec:dustopacities}.  On the other hand,  for optically thick dust ($\tau_\nu\gg1$), \autoref{eq:iblackbody} becomes:
\begin{equation}
I_\nu\sim B_\nu(T_\text{dust})\propto \nu^2
\end{equation}
Thus, the spectral index can be used to infer dust properties only for optically thin dust.  Spectral index $\alpha$ of 3.7 is expected for interstellar medium dust, while millimeter grains or optically thick dust would have  $\alpha\sim 2$. 

{While (sub-)millimeter and centimeter interferometers such as ALMA or VLA are now able to provide high angular resolution observations between $\sim$0.45\,mm and $\sim$1-2\,cm, the above approximations are however only valid over a relatively narrow range of wavelengths. Indeed, if the wavelength is less than roughly 10 times the peak wavelength of the blackbody law at corresponding temperature (Wien's law), the blackbody emission does not follow the Rayleigh-Jeans slope yet. For example, for a temperature of 20\,K typically assumed for the outer disk midplane, the Rayleigh-Jeans approximation is only valid for $\lambda\gtrsim 1\,$mm. On the other hand, at long wavelengths (e.g., centimeter), non thermal processes such as free-free emission or anomalous microwave emission might dominate~\citep[e.g.,][]{Rodmann_2006, Greaves_2022}, limiting the applicability to the above equations to the millimeter range.}\\

Observationally, the integrated spectral index $\alpha$ is often found to be close to 2 in protoplanetary disks, and sometimes even below this limit~{\citep[e.g.,][]{Beckwith_1991, Perez_2012, Tazzari_2021, Garufi_2025}}. Spectral index below two are often found in the inner disks~\citep[e.g.,][]{Dent_2019}.

{Low spectral indices around $\sim2$ could be indicative of the presence of optically thick regions within the disk~\citep{Ricci_2012}. On the other hand, spectral} index below 2 could be explained by two hypothesis. First, if dust is optically thick {and if the dust temperature is sufficiently low, the Rayleigh-Jeans approximation would not be valid. In that case}, the slope of the blackbody emission can be less than 2 at (sub)millimeter wavelength. 
This is because the observing wavelength could be close to the peak of the blackbody emission, and not yet in the powerlaw regime of the blackbody. 
Alternatively, something could be missing in this simple model. 
Indeed, as hinted at in the previous sections (see also Fig.~\ref{fig:opacities}), absorption and scattering opacities of relatively large particles are similar at millimeter wavelengths. Scattering can thus likely not be neglected in the full radial extent of protoplanetary disks. The consideration of this phenomenon can indeed explain spectral index below 2~\citep[e.g.,][]{Zhu_2019, Liu_2019}.

\subsection{Considering scattering}

Without neglecting scattering, and assuming that the disk is azimuthally symmetric, vertically isothermal and infinitely thin, the dust thermal emission from the disk midplane can be computed following~\citep{Sierra_2019}:
\begin{equation}
    I_\nu^\text{abs+scat} = B_\nu(T_\text{dust}) \left[(1-e^{-\tau_\nu/\mu}) + \omega_\nu F(\tau_\nu, \omega_\nu, \mu) \right]
    \label{eq:fulleqscattering}
\end{equation}
where
\begin{equation}
    \begin{aligned}
    \notag
    F(\tau_\nu, \omega_\nu, \mu) = &\frac{1}{e^{-\sqrt{3}\epsilon_\nu\tau_\nu }(\epsilon_\nu -1) - (\epsilon_\nu + 1)}\\
        & \times \Bigg[\frac{1 - e^{-(\sqrt{3}\epsilon_\nu +1/\mu)\tau_\nu}}{\sqrt{3}\epsilon_\nu\mu +1}\\
        &+\frac{e^{-\tau_\nu/\mu} - e^{-\sqrt{3}\epsilon_\nu\tau_\nu}}{\sqrt{3}\epsilon_\nu\mu -1} \Bigg]\\
    \end{aligned}
\end{equation}
and  $\omega_\nu = \kappa^\text{scat}_\nu / ( \kappa^\text{scat}_\nu +  \kappa^\text{abs}_\nu)$ is the albedo, $\epsilon_\nu = \sqrt{1-\omega_\nu}$, $\tau_\nu = \Sigma_\text{dust} \kappa^\text{abs + scat}$ is the optical depth including scattering.
Because \autoref{eq:fulleqscattering} assumes that the disk has a constant temperature along the line of sight and that it is infinitely thin vertically, this equation is only valid for low inclination systems. 

\begin{figure}
    \centering
    \includegraphics[width=\linewidth]{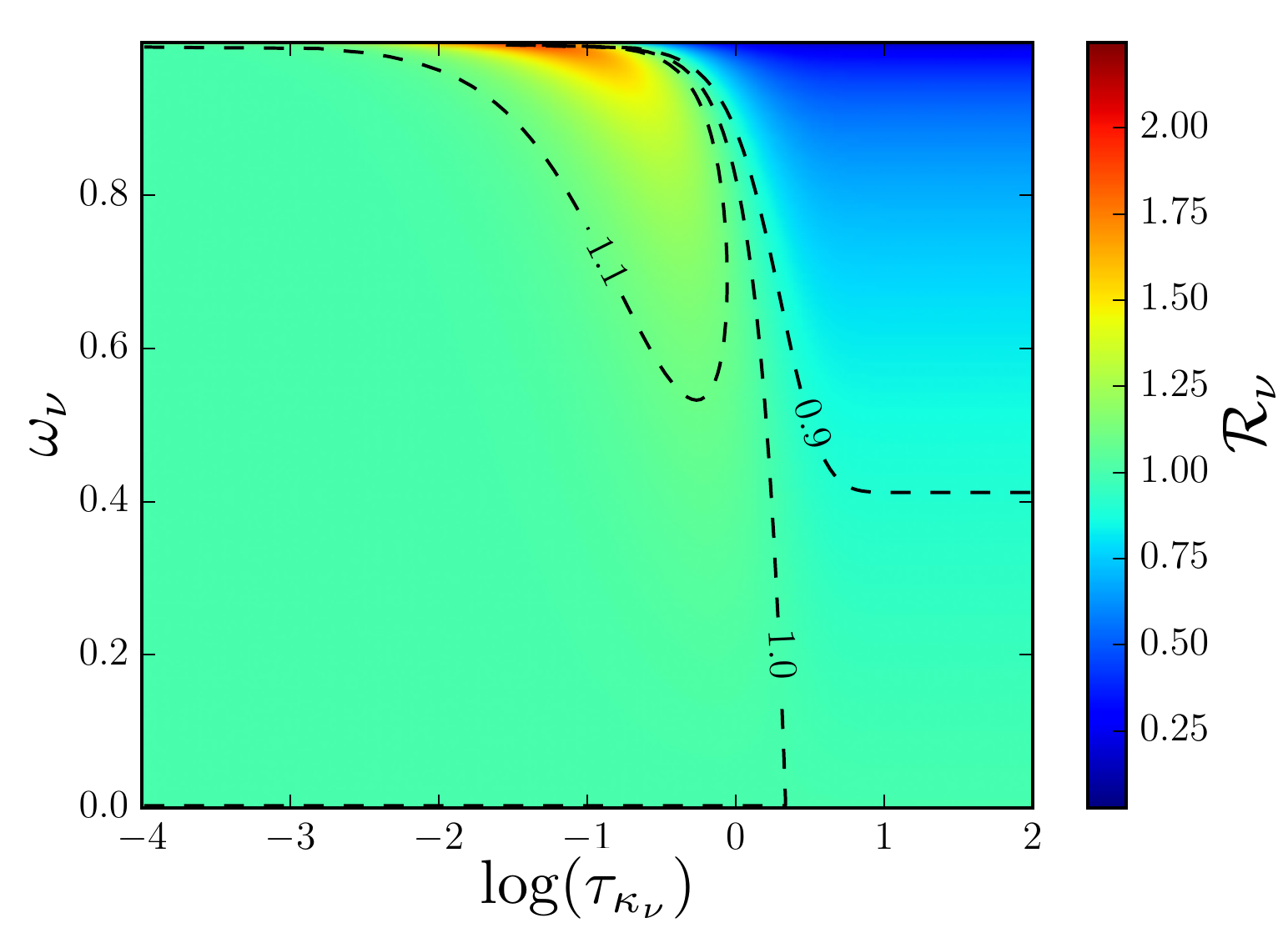}
    \caption{Ratio between the emergent intensity with or without scattering effects, $\mathcal{R} = I_\nu^\text{abs+scat}/I_\nu^\text{abs}$, as a function of the albedo $\omega_\nu$ and the optical depth $\kappa_\nu$, from \cite{Sierra_2020}. Scattering can decrease (top right) or increase (top left) the expected intensity compared to the absorption only case.}
    \label{fig:scattering-absorption}
\end{figure}

When scattering is included, the equation for the expected disk brightness becomes significantly more complex. To better understand the differences implied by considering  scattering, we show in Fig.~\ref{fig:scattering-absorption} the ratio $\mathcal{R} =I_\nu^\text{abs+scat}/I_\nu^\text{abs}$ as a function of the optical depth ($\tau_\nu$) and albedo ($\omega_\nu$). In most of the parameter space, including or not scattering does not significantly change the expected intensity. However, in the case of optically thick dust with a large albedo (top right corner of Fig.~\ref{fig:scattering-absorption}), scattering tends to decrease the expected intensity compared to the case of absorption only~\citep[see also][]{Zhu_2019, Liu_2019}. This is because the mean free path of the particles decreases if scattering is considered compared to the absorption only case, and thus the amount of light that can escape the disk will be less. This can lead to spectral index below 2.  Optically thin dust with high albedo (top left side of Fig.~\ref{fig:scattering-absorption}) can also have an increased intensity compared to the  absorption only case. \\

Recently \autoref{eq:fulleqscattering} has been applied to spatially resolved multi-wavelength observations of protoplanetary disks in the millimeter and sub-millimeter range to infer important disk properties. Specifically, \autoref{eq:fulleqscattering} depends on $T_\text{dust}, \Sigma_\text{dust}$, $\kappa_\nu^\text{abs}$,  $\kappa_\nu^\text{scat}$.  When a dust composition is assumed, the dust opacities are defined by the dust size distribution parameters, namely $a_\text{max}$ and $p$. The later parameter, $p$, describes the powerlaw followed by the size distribution $n(a)da\propto a^{-p}da$. Thus, in principle, observations at 4 different wavelengths allow to retrieve information on the dust properties.

The inclusion of long wavelength observations (e.g., in the centimeter) is particularly critical to obtain robust results. This is because optically thin observations are required to retrieve correct dust properties~\citep{Viscardi_2025}, and  substructures are more likely to be optically thin at long wavelength. High angular resolution observations are also necessary in order to resolve the substructures and avoid beam smearing~\citep{Sierra_2024, Viscardi_2025, Zagaria_2025}. Finally, we note that in very optically thick disk regions, the material detected at different wavelengths might be tracing different disk heights, in which case \autoref{eq:fulleqscattering} is not valid and the analysis can lead to unrealistic grain sizes~\citep{Macias_2021, Guerra-Alvarado_2024}.\\

Most studies using \autoref{eq:fulleqscattering} to infer disk and dust properties find maximum grain sizes between a few centimeters to a few hundreds of microns in protoplanetary disks~\citep{Carrasco-Gonzalez_2019, Mauco_2021, Ueda_2022,  RiviereMarichalar_2024}. {Those studies typically assume compact spherical grains.}  For disks with rings, there are systems which do not show any increase in the dust maximum grain size at the ring location compared to the rest of the disk~(e.g., CQ Tau, SR24S, IM Lup, MWC 480), while other do show larger dust particles within the rings~\citep[HD 169142, GM Aur, AS 209, Elias 24, UX Tau, RX J1615][]{Sierra_2021, Sierra_2024, Carvalho_2024}.  This suggests that some rings are favorable places for grain growth, as we expect in the case of pressure dust traps, while some other might not be. The lack of significant grain growth in some rings could instead indicate that dust sizes are limited by fragmentation~\citep{Jiang_2024}.

The analysis however depends on multiple parameters which have rarely been studied simultaneously. In particular, the assumption on the particle size exponent is degenerate with the maximum grain size. In TW Hya, when $p$ is fixed $a_\text{max}$ tends to decrease with radius, while if $p$ is also allowed to vary then $a_\text{max}$ is nearly uniform while $p$ increases with radius, taking values  between 3.5 and 4.1~\citep{Macias_2021}. 

Porosity also matters in estimating the grain size from the radial SED analysis. Various studies have shown that assuming more porous grains with similar compositions leads to a larger estimate on the value of $a_\text{max}$~\citep{Guerra-Alvarado_2024}. Results suggest that dust in HD\,163296 and CI\,Tau is compact~\citep{Guidi_2022, Zagaria_2025}, while it could be {moderately} porous in HL\,Tau~\citep{Zhang_2023}{, with filling factor around $f\sim0.1$. In the particular case of HL\,Tau, the relatively high millimeter polarization fraction  ($\geq 0.5\%$) can also best be explained by moderately porous dust particles~\citep{Zhang_2023}.}

 Finally, the dust composition affects significantly the analysis described above. Different maximum grain sizes can be obtained using different opacity laws~\citep{Sierra_2025} or local increase at the ring location can be present or not depending on the dust composition assumed~\citep{Guidi_2022}. In a careful analysis of dust composition, \citet{Zagaria_2025} found that the maximum grain size is most significantly affected by the type of organics within the dust, not the amount of water ice present.  Further analysis constraining the best opacity law based on multi-wavelength resolved SED analysis or other type of modeling~\citep[e.g.,][]{Lin_2021, Doi_Kataoka_2023} would be valuable to better understand dust properties and grain growth in disks.

\section{Observational constraints on disk vertical extent and dust settling}
\label{sec:settling}
Observationally characterizing regions of high density in a protoplanetary disk is important to better understand how and where grains grow and planets form. In Sect.~\ref{sec:theorydrift}, we highlighted that because of their interaction with the gas, dust particles are affected by vertical settling and radial drift. Here, we focus on the methods used to obtain observational characterizations of the mechanism of vertical settling.  We refer the reader to the recent reviews of \citet{Rosotti_2023} and \citet{Miotello_2023} for results regarding the radial extent and radial distribution of dust and gas.

Observational claims of vertical settling started with the advent of space observatories like IRAS, 2MASS, Spitzer, and WISE. These missions surveyed large regions of the sky and  enabled the construction of spectral energy distributions (SED) of a significant number of disks, providing insights on their properties. The main predicted effect of vertical settling on the SED is to decrease dust thermal emission for wavelengths larger than $\sim$10$\,\mu$m~\citep{Dullemond_2004, Dalessio_2006}. The maximal impact of settling is encountered at wavelengths around 100\,$\mu$m and is due to the decrease of the emitting height of the dust, which coincides with a decrease of its temperature. Observed SEDs show profiles compatible with the presence of settling, initially identified via Spitzer data between 10 and 30\,$\mu$m~\citep[e.g.,][]{Furlan_2006, Furlan_2011}, and later via radiative transfer modeling of SEDs up to millimeter wavelengths~\citep[e.g.,][]{Grant_2018, Ribas_2020}. This allowed to suggest that settling is active in most Class~II protoplanetary disks. 

SED analysis is, however, usually degenerate with many parameters and does not allow to quantify precisely the efficiency of vertical settling. 
Here we discuss {more direct} observational constraints to the vertical structure of dust in disks obtained from resolved observations in scattered light (optical/near infrared) or at millimeter and centimeter wavelengths. The inclination of the disk is a critical parameter impacting the possible methods available to retrieve its vertical structure. This is why we discuss separately disks viewed edge-on (Sect.~\ref{sec:edgeon}) and those observed at intermediate inclination (Sect.~\ref{sec:intermediateInclination}) in the following sub sections. Face-on systems do not allow for a direct estimate on their vertical extent. 

\subsection{Edge-on disks}
\label{sec:edgeon}
\subsubsection{Small dust and gas}
\label{sec:edgeon_scattered light}
Edge-on and highly inclined systems provide a direct view to their vertical extent, making them ideal targets for studying vertical settling. At optical and near-infrared wavelengths, these disks occult direct stellar emission, and appear as two bright nebulae separated by a dark lane~\citep{Angelo_2023}. The observer primarily detects light from stellar photons, which are scattered by the two disk's surfaces. Because the star is occulted, it is not necessary to use a coronagraph for imaging edge-on disks.

By modeling the scattered light emission of edge-on disks, it is possible to retrieve the scale height of small grains contributing to the disk appearance at these wavelengths. The specific observables that can be used  to do so include the minimum separation of the  two bright nebulae and how the separation of the nebulae varies with distance to the center~\citep[see e.g., review by][]{Watson_2007}. 
Radiative transfer modeling at wavelengths between 0.8\,$\mu$m and 2\,$\mu$m was performed in about half a dozen Class II edge-on disks. These studies typically find small dust scale heights between 7 and 15\,au, at an equivalent radius of 100\,au~\citep{Burrows_1996, Stapelfeldt_1998, Stapelfeldt_2003, Grosso_2003, glauser_2008, Wolff_2017, Berghea_2024}. Younger systems have also been modeled~\citep{Wolf_2003, Sauter_2009, Fischer_2014} but the presence of the envelope complicates the modeling, making the results less robust. At wavelengths up to about 2\,$\mu$m, dust particles are expected to be well mixed with the gas and thus the estimated dust scale heights provide an indirect estimate to the gas scale height. \\

Yet, in some systems, attempts to reproduce observations at multiple optical or near infrared wavelengths using a well-mixed dust model proved challenging, suggesting the presence of vertical dust settling~\citep{Jayawardhana_2002, McCabe_2011}. \emph{JWST} significantly increases the number of systems where resolved images can be obtained at multiple scattering wavelengths. 

Specifically, different studies investigated how the minimum separation of the nebulae varies with wavelengths in different systems. In the case of no significant variation between 2 and 20\,$\mu$m, models indicates that grains with sizes ranging from a few to 10 microns are necessarily present in the upper disk layers. 
On the other hand, significant variation of the darklane thickness with wavelength highlights the absence of grains of a few microns. This is because of the typical shape of the dust opacity laws (see Sect.~\ref{sec:dustopacities}, Fig.~\ref{fig:opacities}). Small grains compared to the observing wavelength show opacities steeply decreasing with wavelength, implying strong variations in the darklane thickness with wavelength. On the other hand, large grains have a uniform opacity at wavelengths smaller than their typical sizes.

Both cases have been observed in Class~II disks. Specifically, \cite{Duchene_2024} and \cite{Tazaki_2024} used  the fact that the darklane thickness does not vary significantly at JWST wavelengths to show that grains up to $3-10\,\mu$m are still high above the midplane in the edge-on systems HH\,30 and Tau\,042021, and that these grains are not significantly affected by vertical settling. This is also consistent with results from \cite{Pontoppidan_2007} using unresolved spectroscopic data on another edge-on disk, the Flying Saucer.  On the other hand, the disk Oph\,163131 shows significant level of settling, with 1\,$\mu$m grains already settled below the $\tau=1$ surface at 4.4\,$\mu$m~\citep{Villenave_2024Oph}. Expanding these studies to a larger number of systems would allow to increase our understanding of the dust distribution and vertical settling efficiency in the upper layers of the disks.\\

In addition, direct molecular gas observations can be used to further directly characterize the gas distribution and properties in edge-on disks.  Using a velocity based tomographic reconstruction, several studies  were indeed able to obtain empiric measurement of the 2D (r, z) temperature profile of edge-on disks~\citep{Dutrey_2017, Teague_2020, Flores_2021}. Such studies can allow to directly characterize the gas vertical extent and properties, and will be  significantly expanded thanks to the \emph{DiskStrat} ALMA large program, which will observe 9 disks with multiple millimeter molecular lines. In addition, the high spectroscopic power of JWST also allows us to probe the spatial distribution of ices~\citep{Sturm_2023, Martinien_2025}, which allows to further constrain disk properties of these systems.

\subsubsection{Dust at millimeter wavelengths}

\begin{figure}
    \centering
    \includegraphics[width=\linewidth, ]{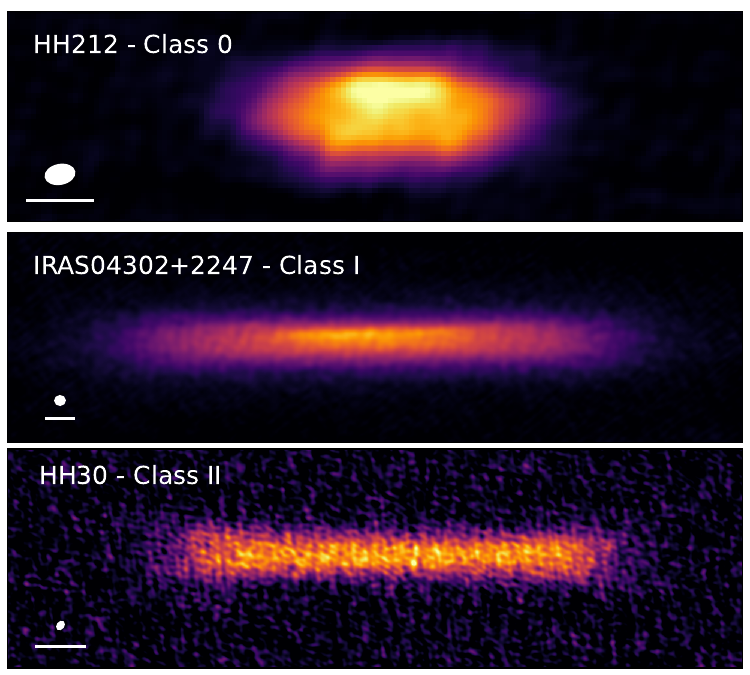}
    \caption{Resolved 1.3\,mm images of edge-on disks at different evolutionary stages~\citep[from][]{Lee_2017, Lin_2023, Tazaki_2019}. The ellipse on the bottom left corner indicates the beam size, while the horizontal bar corresponds to a size of 25\,au.}
    \label{fig:imagesEODmm}
\end{figure}

\begin{figure*}
    \centering
    \includegraphics[width=\linewidth]{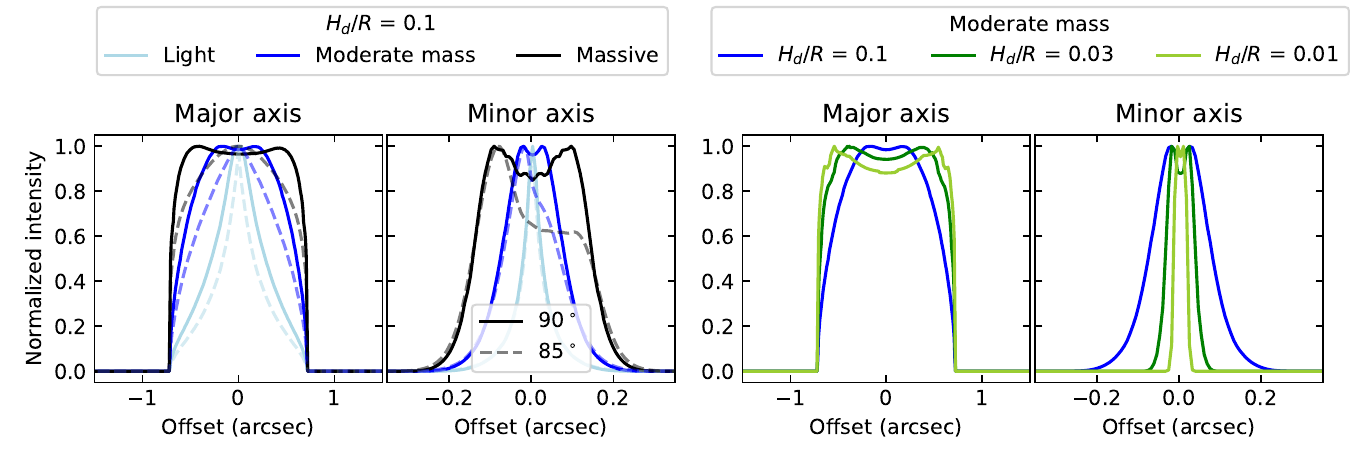}
    \caption{Integrated minor and major profiles of edge-on (solid lines) and highly inclined (dashed lines) disk models for different dust mass (left panels) or different dust scale heights (right panels) computed at 0.89\,mm.}
    \label{fig:EODmm}
\end{figure*}

At the other hand of the spectrum, millimeter and centimeter observations allow us to probe much larger particles, which are expected to be significantly affected by vertical settling. Initial studies of edge-on disks using unresolved millimeter observations, simultaneously modeled with scattered light data, have suggested the need for settling of the bigger grains~\citep{Duchene_2003, Duchene_2010, Grafe_2013, Wolff_2021, Sturm_2023}. However, vertically resolved observations are needed to fully quantify the vertical extent of the larger grains.  

At millimeter wavelengths, edge-on disks typically show a rectangular shape (see Fig.~\ref{fig:imagesEODmm}), where the major axis direction probes the disk radius and the minor axis direction is sensitive to the dust height. It is however important to note here that the apparent vertical height/minor axis extent is not directly equivalent to the dust scale height because of the contribution of different disk regions along the line of sight. To retrieve the vertical scale height of an edge-on disk,  radiative transfer modeling is required, even with observation resolved in the vertical direction. \\

There are however several features which, if present in marginally resolved observations, can point to either a flat or vertically thick dust disk. In Fig.~\ref{fig:EODmm}, we illustrate the impact of 3 main parameters on the appearance of spatially resolved edge-on disks observed at 0.89\,mm: namely the dust scale height, the disk mass, and its inclination.

{To generate Fig.~\ref{fig:EODmm}, we produced models with the radiative transfer code \texttt{mcfost}~\citep{Pinte_2006, Pinte_2009}. We consider a disk around a young sun like star located at 140\,pc with material located between 1\,au and 100\,au. The surface density of the disk follows $\Sigma\propto r^{-1}$, while its vertical extent is characterized by $H_d=H_0(r/100\text{ au})^{1.125}$. We produced models with a dust and gas scale height of $H_0=10$\,au, 3\,au, and 1\,au at the reference radius of 100\,au. The dust mass is also varied between $10^{-5}$\,M$_\odot$, $10^{-4}$\,M$_\odot$, and $10^{-3}$\,M$_\odot$, respectively for the Light, Moderate mass, and Massive models. We assumed a well mixed dust and gas distribution, with particules between 0.03\,$\mu$m and 1\,mm following a typical exponent of $p=3.5$, and generated images at 0.89\,mm for models viewed at an inclination of 90$^\circ$ and 85$^\circ$. The models shown in Fig.~\ref{fig:EODmm} were not convolved by any beam.}  We focus on the size and shape of the integrated major and minor axis profiles, as in previous works~\citep[e.g.,][]{Villenave_2020, Tazaki_2024}.\\

On the leftmost panel of Fig.~\ref{fig:EODmm}, we can see that increasing the disk mass (darker colors) leads the major axis profile of an edge-on disk to transition from centrally peaked to being flat topped with sharp edges. A more massive disk also appears vertically thicker than a lighter source, as we can see from the second panel to the left of the figure. 
Similarly, increasing the dust scale height leads to a larger apparent thickness (rightmost panel). At the same time, increasing the dust scale height also distributes the flux along a larger surface, making the disk globally less optically thick. For the same mass, this implies that the major axis profile will be more centrally peaked and its edges will be less sharp if the scale height is larger (second panel to the right). Thus, the effect of the dust mass and dust scale height is comparable on the minor axis profile but opposite along the major axis profile. Yet, the simultaneous presence of a flat toped major axis profile with a vertically thin minor axis profile in a disk points towards a massive and vertically thin system. 

Observationally, the previously described effects have been used by \cite{Villenave_2020} to conclude that three highly inclined disks showing a flat topped major axis and marginally resolved minor axis have a scale height of a few au only at 100\,au. Compared with typical small dust and gas scale heights of $7-15$\,au at the same radius (see Sect.~\ref{sec:edgeon_scattered light}), this provided some of the first direct evidence for efficient vertical settling in Class~II protoplanetary disks. Later high angular resolution observations of one disk of that sample, HH\,30 (see bottom panel of Fig.~\ref{fig:imagesEODmm}), resolved vertically the disk. Using detailed radiative transfer modeling, \cite{Tazaki_2024} confirmed that significant vertical settling is occurring. \\

Interestingly, in the second panel to the left of Fig.~\ref{fig:EODmm}, we see that for a vertically thick and massive disk the vertical profile does not appear Gaussian. If the disk is perfectly edge-on and the angular resolution sufficient, the minor axis profile shows two bright nebulae of equal intensity separated by a darker region (solid dark line, second left panel). Similarly to scattered light wavelengths, this is because the disk is significantly more optically thick in the midplane than at the disk surfaces. Because of the relative optical depths, dust emission from the midplane is emitted from a larger radius than that detected at higher altitude. At larger radius dust is colder, and thus less bright, leading to the double peaked profile. 

On the other hand, for a disk tilted slightly away from edge-on (85$^\circ$ in Fig.~\ref{fig:EODmm}, dashed dark line), the minor axis profile is asymmetric, with a brighter side and a shoulder. This is because on the near side of the disk, the line of sight crosses more outer disk material than on far side of the disk. On the near side dust becomes optically thick at larger radius than on the back side. Emitting material from the near side is thus colder and its emission is less bright than that of the far side~\citep{Lee_2017, Ohashi_2022}. It is important to note that the millimeter darklane and shoulders are visible only for massive disks which are also sufficiently vertically thick.

Recent observations of young edge-on disks observed at high angular resolution show the ubiquitous presence of such darklane and shoulders. The first evidence of a darklane was obtained in the Class~0 HH\,212 system~\citep[][top panel of Fig.~\ref{fig:imagesEODmm}]{Lee_2017}, for which \cite{Lin_2021} could model observations at millimeter and centimeter wavelengths without requiring any vertical settling. The recent ALMA large program eDISK also revealed asymmetries along the minor axis, similar to the shoulders discussed here, in a large number of young Class~0 or~I highly inclined systems~\citep{Ohashi_2023, Encalada_2024, SantamariaMiranda_2024, Takakuwa_2024, vanthoff_2023}. This suggests that these disks are not or very modestly affected by vertical settling. This is indeed consistent with results based on more detailed radiative transfer models of vertically resolved observations of the shoulder bearing disk IRAS\,04302 at multiple millimeter and centimeter wavelengths~\citep[][middle  panel of Fig.~\ref{fig:imagesEODmm}]{Lin_2023, Villenave_2023}. Other observations and modeling of marginally resolved young systems suggested that they are vertically thick at millimeter wavelenghts~\citep{Zhang_2020, Michel_2022, Sheehan_2022, Ohashi_2022}, suggesting that settling might not be efficient in the Class~0 and~I phase.

However, we note that constraints on the true gas scale height are currently limited in edge-on systems, as they have been mostly inferred from scattered light dust observations, which is particularly difficult for young systems with an envelope. Several constraints on settling rely on gas scale height estimates based on the midplane temperature of a radiative transfer model (\autoref{eq:gasScaleHeight}), which will need to be improved in the future with direct gas constraints.

\subsection{Intermediate inclination disks}
\label{sec:intermediateInclination}
Intermediate inclination systems constitute the majority of known disks. Thanks to observational large programs, such as DSHARP with ALMA~\citep{Andrews_2018}, DARTTS~\citep{Avenhaus_2018}, DESTINYS~\citep{Ginski_2024, Garufi_2024, Valegrard_2024} with VLT/SPHERE, or GEMINI-LIGHTS~\citep{Rich_2021} with GPI, many inclined protoplanetary disks have been observed at high angular resolution. These studies revealed that substructures are ubiquitous in protoplanetary disks in the Class~II phase, with gaps and rings being the most common features~\citep[e.g.,][]{Huang_2018, Long_2018, Andrews_2020, Bae_2023}. Here we highlight how the vertical extent of dust grains and gas can be characterized via spatially resolved observations, providing us constraints to vertical settling.

Currently, as for edge-on systems, directly estimating the gas or dust scale height from observations is difficult. This is because small dust and bright molecular lines are highly optically thick ($\tau_\nu\gg1)$. In that case, as highlighted in \autoref{eq:iblackbody} and Sect.~\ref{sec:classical_analysis}, the intensity is not sensitive to the disk surface density $\Sigma$. 
Thus, the apparent vertical extent of the disk with these tracers does not correspond to their respective dust scale height, but corresponds instead to the height of their $\tau=1$ surface, which can be several times higher than the density scale height. In simple words, the  $\tau=1$ surface corresponds to the lowest surface where a photon can emerge from the disk. For small dust, this corresponds to the surface of the last scattering event.

Detailed thermo-chemical or radiative transfer modeling of individual sources can be used to estimate the scale height of intermediate inclination systems. However, the results are model dependent and time consuming to obtain and have thus been performed only to a limited number of sources~(e.g., \citealt{Muro-Arena_2018, Villenave_2019, Martinien_2024} for scattered light models or \citealt{Calahan_2021, Kamp_2023} for thermochemical estimates). 

Here we highlight geometrical methods that allow to determine the disk surface from molecular line emission, scattered light observations, and millimeter dust emission. Although they do not trace directly the dust and gas scale heights they are indicative of the optical thickness of the different tracers and could later be used for estimating the small dust or gas scale heights.

\subsubsection{Molecular gas emission}
\label{sec:moleculargasemission}

For optically thick gas emission lines, it is possible to infer the height of the emitting layers using a geometrical method which does not require the presence of substructures. Introduced by \cite{Pinte_2018}, the method requires spectrally and spatially resolved observations where the near and far sides of the disk can be disentangled in a single channel map. Such observations can be obtained with instruments like ALMA for example. While the method also allows to infer the local gas velocity~\citep{Pinte_2018, Teague_2019}, we focus here on the height of the emission only. 

\begin{figure}
    \centering
    \includegraphics[width=\linewidth]{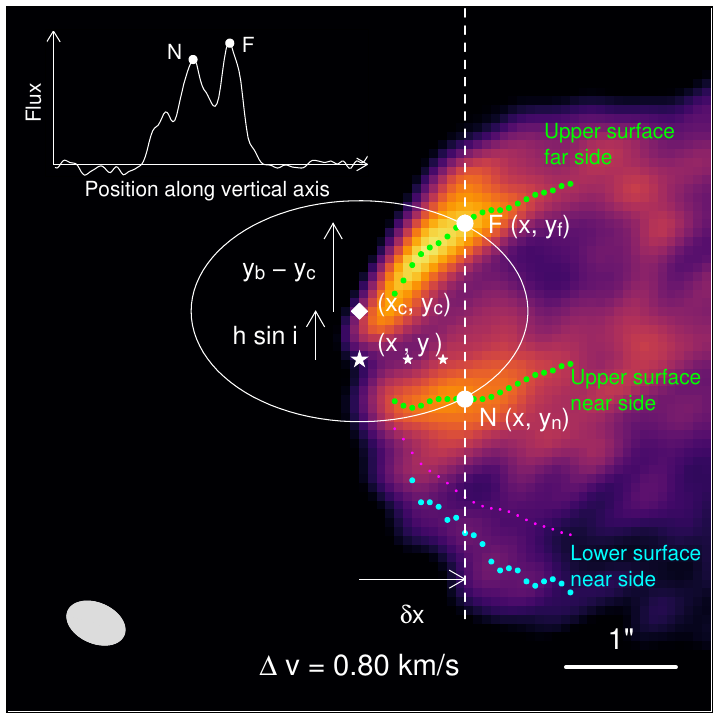}
    \caption{Single channel map with a schematic of the quantities used to estimate the height of the emitting surface. The central object is marked by a star. For every $x$ along the major axis direction, the minor axis maxima can be identified. Connecting these points by an ellipse of a unique height and centered on the star it is possible to retrieve the height of the emitting layer~(\autoref{eq:COheight}). Credit: \citealt{Pinte_2018}, A\&A, 609, A47, reproduced with permission \textcopyright\ ESO }
    \label{fig:heightmolecules}
\end{figure}

The gas is assumed to follow a circular orbit and to be well described by Keplerian rotation.  Radial and vertical gas motions are thus neglected here. Then, for every $x$ along the major axis direction, one can identify the peaks of the channel maps along the minor axis direction, corresponding to the near and far disk surface, $y_\text{n, near side}$ and $y_\text{f, far side}$. This is illustrated in Fig.~\ref{fig:heightmolecules}. These points can be connected by an ellipse of a given height above the disk plane, where the center of the corresponding projected circular orbit have coordinates $x_\text{c} = x_\star$ and $y_\text{c} = (y_\text{n, near side} + y_\text{f, far side})/2$.
Knowing the disk inclination, $i$, one can retrieve the radius, $r$, and height, $h$, of the emission following:
\begin{equation}
     \left\{
    \begin{aligned}
      & r = \sqrt{(x - x_\star)^2 + \left(\frac{y_\text{f, far side} - y_\text{n, near side}}{\cos i}\right)^2 }\\
      & h = \frac{y_\text{c} - y_\star}{\sin i}
    \end{aligned}
    \right.
    \label{eq:COheight}
\end{equation}

Numerical implementations of this method are publicly available in softwares such as \texttt{disksurf}~\citep{Teague_2019}, \texttt{DISCMINER}~\citep{Izquierdo_2021, Izquierdo_2023}, or \texttt{ALFAHOR}~\citep{PanequeCarreno_2023}.\\

This method has first been applied to CO isotopologues emission, which are brighter and thus easier to detect than other molecules. The results show a stratification of CO isotopologues, with $^{12}$CO tracing a higher altitude than $^{13}$CO, which is itself higher than C$^{18}$O~\citep{Pinte_2018, PanequeCarreno_2021, Law_2023, Galloway_2025}. \cite{Law_2021} reports $^{12}$CO surfaces with $h/r$ between 0.2 and 0.5, while $^{13}$CO typically reached $h/r<0.2$, and C$^{18}$O  is constrained to $h/r<0.1$ for the disks of the MAPS large program~\citep{Oberg_2021}. 

When the same molecular tracer is analyzed (e.g., $^{12}$CO), some diversity in the emission heights is also found between sources. Specifically, the emission height is strongly correlated with the outer disk size, while it is only weakly or tentatively correlated to the stellar mass~\citep{Law_2022, Galloway_2025}. 
More recently, rarer molecules have also been studied to estimate their emission heights in individual systems, such as CN, HCN, H$_2$CO, HCO$^+$, C$_2$H, c-C$_3$H$_2$, CI, C$_2$H$_2$, HCO$^+$~\citep{PanequeCarreno_2022, PanequeCarreno_2023, HernanderVera_2024, Law_2023CI, Urbina_2024, PanequeCarreno_2024, Law_2024}. These studies map the vertical stratification of molecules and allow us to obtain empirical estimates of the temperature profile of protoplanetary disks, which are, so far, broadly consistent with expectations from radiative transfer modeling.\\

To go beyond empirical measurement of the molecular emission height, \cite{PanequeCarreno_2023} proposed a method to estimate the gas scale height from the height of the molecular emission, assuming a vertical density profile distribution and a good knowledge on the surface density profile~\citep[see also][for similar analysis inferring surface density profile based on an assumption on the gas scale height]{Rosotti_2025}. The method relies on the assumption that $^{12}$CO emission traces the region between the point where the emission becomes optically thick and that where CO becomes self-shielding. By integrating the assumed vertical density profile along these boundaries it is possible to estimate the gas scale height. Implemented on 5 disks, their results show gas scale height of about $h/r\sim0.1$ beyond about 100\,au, which implies that the $^{12}$CO emitting layers are located between 2 and 5 times above the disk scale height. 

Alternatively, similarly to edge-on disks, estimates of the gas scale height based on \autoref{eq:gasScaleHeight} and assuming a midplane temperature based on the stellar luminosity can also be compared to the emitting height levels, but are more uncertain.  In the future, further developments on the comparison of apparent emission heights and true gas scale height are necessary to better understand disk structure and  to further characterize vertical dust accumulation in the disk midplane.

\subsubsection{Dust in the optical to near infrared}

At optical to near infrared wavelengths, dust is expected to be well mixed with the gas. Due to the high optical depth, the scattering surface is expected to be located high above the disk midplane. In that context, an inclined ring with a non negligible height will appear as an ellipse which is off-centered from the central source along the minor axis direction~\citep[e.g.,][]{deBoer_2016, Ginski_2016, Monnier_2017}. This is illustrated on the left panel of Figure~\ref{fig:sketchheight}.

Measuring this offset along the minor axis, $\text{offset}_\text{minor}$, can then allow us to determine the height of its $\tau=1$ surface, $h_{\tau=1}$. To perform this analysis, rings are generally assumed to be perfectly circular and axisymmetric. 
Disks and ring parameters such as the ring radius, $R_\text{ring}$, the inclination, $i$, and the position angle must also be known precisely. They can for example be derived from an ellipse analysis. Knowing these parameters allows us to retrieve the scattering height following~\citep{Avenhaus_2018}:
\begin{equation}
    \text{offset}_\text{minor} =  R_\text{ring} \frac{h_{\tau=1}}{r}\sin(i)
\end{equation}

This analysis has been performed for about a dozen of sources observed in J (1.245$\mu$m), H (1.625$\mu$m) and K (2.182$\mu$m) bands~\citep{Avenhaus_2018, Ginski_2024, Derkink_2024, Roumesy_2025}. {A numerical implementation of this method is publicly available in the \texttt{DRAGyS} software~\citep{Roumesy_2025}.}
This geometrical method typically finds values of $h_{\tau=1}/r$ between 0.1 and 0.35, with a median aspect ratio around 0.2. The aspect ratio can be up to 3 times larger than the typically expected gas scale height  aspect ratio of 0.1. 
No clear trend is found on the dependency of the scattering height with observing wavelength or with stellar mass. However, within one system, the ratio $h_{\tau=1}/r$ tends to increase with radius, suggesting that the last scattering surface of disks is flared.

\begin{figure}
    \centering
    \includegraphics[width=\linewidth]{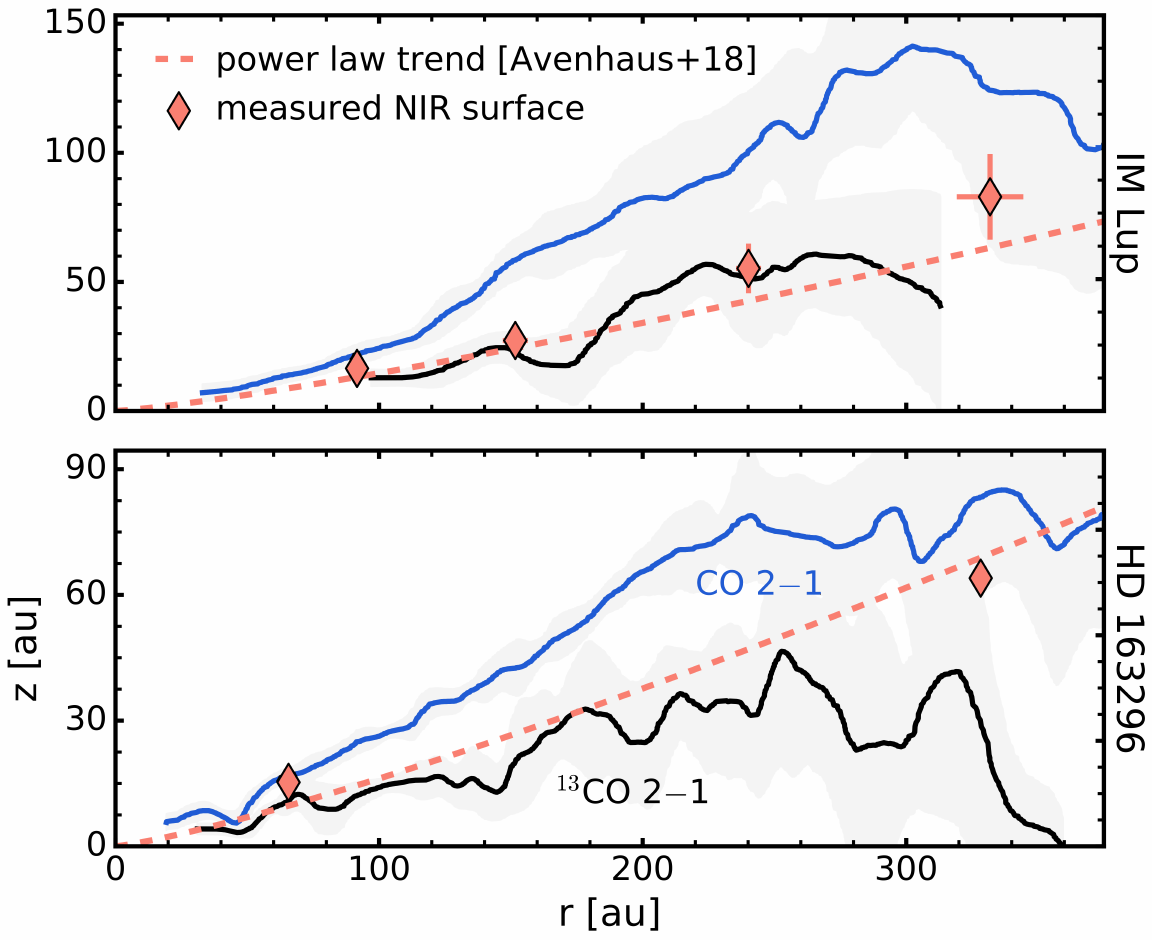}
    \caption{Emission surfaces of $^{12}$CO (blue) and $^{13}$CO (black) in two well characterized disks (adapted from \citealt{Law_2021}). The diamond show $h_{\tau=1}$ for individual rings based on scattered light observations. The red dashed line shows inferred scattered light surfaces for a sample of disks. }
    \label{fig:comparisonheights}
\end{figure}

\begin{figure*}
    \centering
    \includegraphics[width=\linewidth]{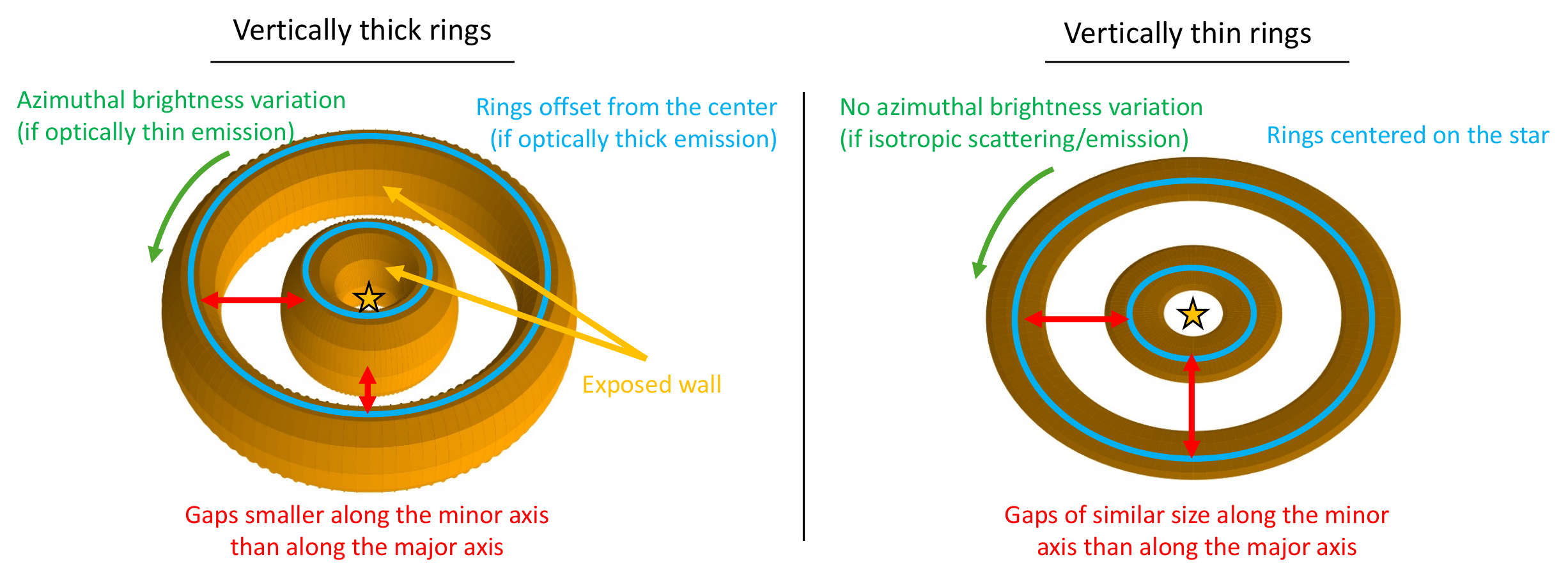}
    \caption{Schematic representation of different characteristics expected in a vertically thick (left) or vertically thin (right) disk, with rings.}
    \label{fig:sketchheight}
\end{figure*}

Comparing the CO isotopologue emission height (see Sect.~\ref{sec:moleculargasemission}) with the scattering surfaces inferred from dust observations can yield a variety of results depending on the source. Fig.~\ref{fig:comparisonheights} shows the $^{12}$CO and $^{13}$CO emission height compared with scattered light heights for two well characterized disks. The figure shows that $^{12}$CO and scattered light emission may originate from a similar layer at short radius ($R\lesssim100$\,au) while at larger distance from the star $^{12}$CO is typically emitted from a region of higher altitude~\citep{Rich_2021, Law_2021}. However a large variety of behavior are found between sources which makes it hard to devise a generic conclusion. Methods able to infer the dust scale height from the scattered light emitting surface would allow to significantly expand such constraints on low inclination systems.

\subsubsection{Dust at millimeter wavelengths}

At millimeter wavelengths, dust is typically optically thinner than in the optical or infrared, which allows the observer to trace material closer to the midplane. Thus, rings are typically not strongly offset from the central location at these wavelengths and instead other geometrical effects can be used to infer the vertical extent of the disk structures. Fig.~\ref{fig:sketchheight} summarizes the main effects allowing to estimate the dust scale height at millimeter wavelengths (see also \citealt{Villenave_2025}). We note that all of the following methods require sufficiently high angular resolution to resolve the substructures and identifying the detailed effects. The rest of this section presents the different effects expected from vertically thin or thick rings and some of the constraints that were obtained using the different methods. \\

One of the most recent effect that has been identified as relevant at millimeter wavelengths is the so-called \emph{wall effect}~\citep{Ribas_2024}, which has been well known from scattered light observations~\citep[e.g.,][]{Duchene_2004, Krist_2005}. For optically thick dust with a significant vertical extension, the far side of the disk (wall) is directly illuminated by the central object and also directly visible by the observer. On the other hand, the illuminated near side of the disk is not directly visible by the observer, and thus appears less bright. The wall effect leads to brightness asymmetries along the minor axis profile, with the far side being brighter than the near side. 

This effect has been identified and modeled in a few disks, allowing to estimate low dust scale height aspect ratio $h_{mm}/r$ of about 0.01 and 0.02, respectively at 37\,au in CIDA\,9 and at 12\,au in RY\,Tau~\citep{Ribas_2024}, and larger than 0.08 in HL\,Tau at about 32\,au~\citep{Guerra-Alvarado_2024}. Interestingly, \cite{Ribas_2024} identified 10 potential additional candidates for the wall effect, for which the modeling could help obtaining a statistical view of the vertical structures relatively close to the star. \\

Besides the potential presence of a wall, the vertical thickness of rings or of the outer disk leads to other measurable effects on the minor axis profiles. Specifically, as illustrated in Fig.~\ref{fig:sketchheight}, the depth and apparent width of a gap is expected to vary with azimuth in the case of a geometrically thick disk. A thick disk will have less pronounced gaps along the minor axis direction than along the major axis direction, or than a thinner disk. Such effects are detectable at a gap location but also at the outer edge of a disk. In that case a thinner disk will have a steeper outer edge profile along the minor axis direction than a thicker system. 

Since its introduction by \cite{Pinte_2016} on HL\,Tau, this method has been used on nearly two dozen systems~(\citealt{Villenave_2022, Pizzati_2023, Villenave_2025, Antilen_in_prep}) allowing to place constraints to the height of the outer disk at several radial locations. In most cases, this method only allowed to place upper limits to the allowed dust scale heights, revealing that the outer regions of Class~II protoplanetary disks are significantly affected by vertical  settling. Most systems are found to have dust aspect ratio $h_{mm}/r$ lower than 0.05 and even as low as 0.005. These constraints necessitate sufficient resolution to resolve the gap structures (typically $<0.1''$ for disks within $\sim140$\,pc) and are thus mostly sensitive to the outer disk regions ($R\gtrsim50$\,au). \\

In the case of narrow rings, a third effect has been pointed out for disks by \cite{Doi_Kataoka_2021}. Optically thin, vertically thick, but radially narrow rings are expected to show azimuthal brightness variations. This is because of the variation of the amount of material along the line of sight as a function of the azimuth. Along the minor axis direction the projected column density is reduced compared to its value along the major axis direction. As a result, the ring is predicted to be brighter along the major axis direction than along the minor axis direction, if dust is optically thin and the ring vertically extended. However, caution is required because an elongated beam could mimic this effect. Radiative transfer modeling, with synthetic observations with similar observational parameters are necessary to fully conclude on the presence of the ring effect or not. 

Used in about half a dozen disks, this method allowed to identify vertically thick rings in the disks of HD\,163296, LkCa\,15, and V1094\,Sco~\citep{Doi_Kataoka_2021, Liu_2022, Doi_Kataoka_2023, Villenave_2025, Jiang_2025}, with $h_\text{mm}/r>0.05$ respectively at 67, 69, and 134\,au. 
These results are particularly interesting for several reasons. First, the vertical extent of the dust in the inner ring of these sources contrast with the low upper limits found at similar distance to the central star in other systems, which are typically significantly lower.  Second, at larger distances, these three systems show vertically thin outer disk, which indicates a decrease in the ratio $\alpha_z/St$ with radius (see Sect.~\ref{sec:theorydrift}, \autoref{eq:alphaZ_St}). In addition, in the context of core growth by pebble accretion, thinner disks are also more favorable for forming planets than thicker systems. Thus, in HD\,163296, LkCa\,15, and V1094\,Sco, planet formation might be more efficient in the outer disk ($R>100$\,au) than in the inner ring ($R\sim 50$\,au), strongly modifying our view of planet formation. \\

Finally, we note that another type of azimuthal brightness variation is expected in the presence unresolved optically thick rings following the azimuthal direction of the disk. Mechanism such as the streaming instability can generate such structure of sub-au scales. \cite{Scardoni_2024} showed that in the presence of such unresolved optically thick dust rings, the minor axis profile of a disk is expected to be brighter than the major axis profile. This is opposite to the behavior described for the ring effect, which is expected in the case of optically thin rings. This effect has not been identified in observations yet. While it would likely not be relevant to estimate the vertical extent of the substructures, identifying such effect would reveal the presence of unresolved dense substructures, which are presumably favorable spots for grain growth and/or planet formation.\\

To summarize, at millimeter wavelengths, the disks are found to have low dust scale heights, suggestive of vertical settling. Constraints on the radial variation of disk thickness are rare and reveal that a few systems have a thicker inner ring compared to the outer disk (at locations around 50 vs 100\,au). Those systems might not be common and further analysis are necessary to identify the vertical dust structure at different radius.

\section{Summary {and outlook}}
\label{sec:conclusion}

This tutorial aimed to cover basic predictions for dust transport and evolution and link them with current constraints on dust and gas vertical extent in protoplanetary disks.  
Observations at a wide range of wavelengths, in the optical/near infrared and millimeter/centimeter range, are now able to obtain spatially and spectrally resolved images of protoplanetary disks. This provides us with unique opportunities to get direct insights towards the spatial distribution of their different components. 
We focused on detailing the methodologies employed to infer dust grain size from multi-millimeter range analysis, and on estimating dust height of edge-on to moderately inclined disks using various different tracers, in order to provide an introduction for beginner. Geometrical methods and radiative transfer modeling are essential and complementary tools to characterize the density structure of disks.

Some of the main observational results are that grains of a few millimeter in size are present in disks. Grains are not necessarily larger in rings than in other regions of the disks, depending on the system. Moreover, millimeter dust grains are often found to be affected by vertical settling in the outer regions of Class~II protoplanetary disks, while settling might be less efficient in younger systems. However, improved methods or modeling works are necessary to constrain the gas scale height in a larger number of systems, in order for the dust constraints to be quantitatively compared and infer precise constraints on vertical settling. 

{
The analysis of the local grain size distribution allows us to study the implications for grain and planet growth efficiency. One challenge is that current estimates of the particle size exponent $p$ remain rare~\citep{Macias_2021, Doi_Kataoka_2021}, and prevent us from a direct confrontation with predictions from different growth barriers (fragmentation, bouncing, radial drift, see Sect.~\ref{sec:theorydrift}). However, the moderate maximum grain size typically found in systems with rings and gaps suggests that dust growth might be limited by fragmentation in rings~\citep{Jiang_2024}. In that case, disk turbulence needs to be low and grains must be fragile, as suggested by experiments and numerical simulations. Assuming dust size is limited by fragmentation or assuming a value for the gas to dust ratio, different studies typically find that grains detected at millimeter wavelength have Stokes number in the range of $10^{-3}-1$~\citep[e.g.,][]{Sierra_2019, Zagaria_2023, Doi_Kataoka_2023, Jiang_2024, Villenave_2025}.  } 

{Beyond the initial growth of dust particles, the estimate of the local maximum dust size can be used to estimate the potential for gravitational instability of a disk region~\citep[e.g.,][]{Sierra_2021} and thus for the direct formation of planetary cores. When this analysis is performed, very few systems appear to be consistent with gravitational instability. Notable exceptions are Elias\,27 and AB\,Auriga, which also show large scale spiral arms consistent with prediction from gravitational instability~\citep{PanequeCarreno_2021, Speedie_2024}.}

{However, besides the gravitational instability other mechanisms such as the streaming instability and efficient pebble accretion might also be at play to form planets in disks. Interestingly, the vertically settled outer disks appear as favorable locations for efficient core growth by pebble accretion~\citep{Villenave_2022, Villenave_2025}. In such systems, most of the pebble flux could be intercepted by a planetary core, allowing for an efficient growth~\citep{Lambrechts_2019}. On other hand, radially narrow and vertically thin rings, such as the outer ring of HD163296, might be dense enough to trigger the streaming instability and quickly form planetesimals~\citep{Zagaria_2023}. }

{To further advance our understanding of planet formation, the analysis described above need to be performed on a larger, statistically significant sample of disks. Higher angular resolution observations at multiple wavelengths are critically needed to better understand the dust size distribution, their composition, and porosity levels. Such observations would also be used to perform direct search for on-going streaming instability, for example via the azimuthal brightness profiles of rings~\citep{Scardoni_2024}, and further study the inner regions of protoplanetary disks for which the grain and planetesimal growth potential remain relatively unexplored. 
Extending studies to longer wavelengths, with next-generation observatories like the Square Kilometre Array (SKA) and the next-generation Very Large Array (ngVLA), will also be critical to characterize larger dust particles, which  are fundamental building blocks for planet formation. The next frontier to fully link the dust properties and their dynamics also requires a better understanding of the gas distribution, composition, and mass, which is currently being tackled by several ALMA large programs. Finally, direct detections and characterization of protoplanets embedded in disks, by current or future instrumentation such as JWST or the ELT, would strongly benefit the field and improve our understanding on the processes leading to their diversity. }\\

\medskip

\emph{Acknowledgements:} {MV thanks the referee for their careful reading of the manuscript and suggestions for critical additions which improved the quality of the document.}
I would like to thank the disk group in Grenoble around the ERC Dust2Planets (PI: Fran\c cois M\'enard) for the interesting discussions during the writing process and especially Maxime Roumesy, Laurine Martinien, {and} Karl Stapelfeldt for critical reviews of the manuscript. I also thank the editor for the invitation to write this tutorial{, Myriam Benisty for her recommendation, and} acknowledge support from CNRS INSU. This paper makes use of the following ALMA data:  2017.1.00712.S, 2019.1.00261.L, and  2017.1.01701.S.

\bibliography{biblio}{}
\bibliographystyle{aasjournal}

\end{document}